\documentclass[fleqn,usenatbib]{mnras}
\usepackage[T1]{fontenc}
\usepackage{ae,aecompl}
\usepackage{graphicx}   
\usepackage{amsmath}    
\usepackage{amssymb}    
\usepackage{float}

\usepackage[dvipsnames]{xcolor}

\graphicspath{{figures/}}

\title[Core shift in parabolic accelerating jets]%
{Core shift in parabolic accelerating jets}
\author[Nokhrina E.E.]{\parbox{\textwidth}{
E.~E.~Nokhrina$^{1,2}$\thanks{E-mail: nokhrinaelena@gmail.com},
A.~B.~Pushkarev$^{3,1}$
}
\vspace{0.4cm}\\
\parbox{\textwidth}{
$^1$Lebedev Physical Institute, Leninsky prosp.~53, Moscow, 119991, Russia\\
$^2$Moscow Institute of Physics and Technology, Dolgoprudny, Institutsky per., 9, Moscow region, 141700, Russia\\
$^3$Crimean Astrophysical Observatory, Nauchny 298688, Crimea, Russia
}
}

\begin{document}

\date{Accepted 2024 January 11. Received 2024 January 11; in original form 2023 September 28}

\pagerange{\pageref{firstpage}--\pageref{lastpage}} \pubyear{}

\maketitle

\label{firstpage}

\begin{abstract}
The core shift method is a powerful method to estimate the physical parameters in relativistic jets from active galactic nuclei. The classical approach assumes a conical geometry of a jet and a constant plasma speed. However, recent observations showed that neither may hold close to the central engine, where plasma in a jet is effectively accelerating, and the jet geometry is quasi-parabolic. We modify the classical core shift method to account for these jet properties. We show that the core shift index may assume values in the range $0.8-1.2$ or $0.53-0.8$ depending on the jet geometry and viewing angle, with the indices close to both values are indeed being observed. We obtain the expressions to estimate jet magnetic field and a total magnetic flux in a jet. We show that the obtained magnetic field value can be easily recalculated down to the gravitational radius scales. For M~87 and NGC~315 these values are in good agreement with the ones obtained by different methods.

\end{abstract}


\section{Introduction}
\label{s:intro}

It is now a generally accepted model that the ordered magnetic field in the vicinity of the supermassive black holes (SMBH), residing in the centres of active galactic nuclei (AGN), play the key role in launching relativistic jets \citep{BMR-19}. It is the presence of a magnetic field that either threads the black hole ergosphere and establishes the Blandford--Znajek process \citep{BZ-77}, responsible for the extraction of a BH rotational energy to supply the jet power \citep{Tchekhovskoy_11}, or launches a jet from a disc via Blandford--Payne mechanism \citep{BlandPayn82}. The Poynting flux at the jet base defines the maximum Lorentz factor, that can be achieved if all the energy of electromagnetic field is transformed into a plasma bulk motion kinetic energy \citep[e.g.,][]{Beskin10}.  

Many efforts have been done to understand the magnitude and geometry of a jet magnetic field at different scales. Polarisation measurements in radio band yields us information about possible B-field geometry. Multiple studies at parsec scales reveal the helical magnetic field expected from the theoretical modelling \citep{Gabuzda17, Gabuzda18}. The unprecedented radio images of M87 allowed to confirm that the helical magnetic field B is maintained at the scales from $300$~pc up to $1$~kpc \citep{Pasetto21}. The observed Faraday rotation measure (RM) changing sign can be explained by the presence of toroidal magnetic field around a jet in 3C 273 up to the distances of order of $500$~pc de-projected \citep{Lisakov21}. The detection of a large scale toroidal magnetic field at the kpc scales \citep{Knuttel17} indicates the strong electric current flowing in the relativistic jets far from the expected acceleration and collimation domain \citep{BMR-19, Kov-20}, showing the importance of a Poynting flux even at such distances from the central engine. Direct polarimetry of the event horizon scales in M87 jet by \citet{EHT_B} provides the B-field magnitude of the order of a few to tens Gauss.   

One of the most robust methods to estimate the magnetic field amplitude on the observational basis is using the core shift measurements. The core shift effect is observed in jets due to the changing physical properties (flow velocity, magnetic field and emitting plasma number density) along an outflow, so that the surface with optical depth equal to unity appears at the different positions for the different observational frequencies \citep{L98, hirotani2005}. Together with the model assumptions, core shift measurements provide for the magnetic field typical values of the order of $1$~G at $1$~pc distance \citep{L98}. The results obtained using this method are in general agreement with other expected jet parameters, such as initial magnetisation and jet composition \citep{hirotani2005, Sullivan09_coreshift, Zdziarski-15, NBKZ15, Pushkarev12}. 

The core shift method as its important ingredients uses the following assumptions: equipartition between the proper frame emitting particles and magnetic field energy density,
certain scalings with the distance of a magnetic field $B$ 
and plasma number density $n$ \citep{BlandfordKoenigl1979} and conical jet geometry. The observed jet boundary shape --- the dependence of a jet width $d$ on a distance along a jet $r$ --- can be approximated by the power-law
\begin{equation}
d=a_1r^{k}. \label{d_r}
\end{equation}
The extensive study of jet shapes \citep[see e.g.][and references within]{MOJAVE_XIV} reveals in many sources approximately constant power $k\approx 1$, corresponding to a conical or quasi-conical geometry with index $k$ being close to unity. However, recent studies of a dozen of nearby jets demonstrate, that their shapes are consistent with two power laws, smoothly transiting from one to another on more or less short scale of  the overall length range \citep{Asada12, tseng16, Hada18, r:Akiyama18, r:Nakahara18, r:Nakahara19, r:Nakahara20, Kov-20, Park_2021, Boccardi_2021, Okino22}. For these sources the typical values of $k$ closer to the jet launching point is $k\approx 0.39-0.59$. This we call quasi-parabolic, or parabolic for short. Further downstream the shape becomes closer to conical (quasi-conical) with $k\approx 0.73-1.4$ \citep[e.g.][]{Asada12, Boccardi_2021, Kov-20}. The implicit study of a geometry of the more distant sources shows that the cores at 15 and 8 GHz belong to a quasi-parabolic part of a jet \citep{Nokhrina2022}.
This means that one may expect the radio cores observed at 
high frequencies $230-15$~GHz and, may be, down to $8$~GHz, being in a parabolic domain and with a plasma being still accelerated.
This may impact the results by classical core shift method, which uses typical observational frequencies $8$--$15$~GHz \citep{Pushkarev12}. 

Parabolic accelerating jet model alters the $k_\mathrm{r}$-index in the frequency dependence of the core position $r_\mathrm{core}\propto \nu^{-1/k_\mathrm{r}}$ \citep{Ricci22}. The multi frequency observations for M87 by \citet{Hada11} and for 20 sources by \citet{Sokolovsky11} demonstrate that $k_\mathrm{r}$ is indeed close to unity. This may be explained by either the conical jet model with a constant Lorentz factor $\Gamma$, which may be the case for the sample by \citet{Sokolovsky11}, or the quasi-parabolic accelerating jet model observed at the maximum Doppler factor \citep{Ricci22} for M87 jet with a clear quasi-parabolic shape. On the other hand, a dedicated search in NGC 315 by \citet{Ricci22} yields the change in $k_\mathrm{r}$ index from 0.57 to 0.93 approximately at the jet shape transition point. 
Using different methods, \citet{kutkin_etal14} found the exponent $k_\mathrm{r}$ in range 0.6–0.8 for quasar 3C 454.3. The typical $k_\mathrm{r}$-indices for the MOJAVE sample is estimated to be approximately $0.83$ by Kravchenko et al. (in preparation).

Accounting for parabolic accelerating jets alters the scalings of a magnetic field $B$ and plasma number density $n$ as functions of a distance from the central source $r$, introduces non-constant plasma bulk flow motion Lorentz factor $\Gamma$ and  changes the way of extrapolating the magnetic field amplitude down to the gravitational radius scales \citep{Ricci22}. In this paper we modify the core shift method for parabolic jet boundary geometry and estimate the magnetic filed amplitude at the gravitational radius for ten sources.

The paper is organised as follows. In Section~\ref{s:core_shift_parab} we present the model for magnetic field and emitting particles number density to obtain the possible core shift indices $k_\mathrm{r}$ for a parabolic accelerating jets. In Section~\ref{s:core_shift_B} we discuss the relation between magnetic field and plasma number density and obtain the expressions to estimate both values. In Section~\ref{s:Bfield} we review the data to estimate $B$. We discuss our results and methods in Section~\ref{s:discussion}. Finally, we summarise our work in Section~\ref{s:summary}.

\section{Frequency dependent apparent core shift in parabolic accelerating flow}
\label{s:core_shift_parab}

Emission and opacity of a jet results from the emitting plasma population with assumed power-law energy distribution in the plasma proper frame, designated by the asterisk:
\begin{equation}
\mathrm{d}n_\mathrm{syn*}=K_{e*}\gamma^{-1-2\alpha}\mathrm{d}\gamma,\quad \gamma\in[\gamma_\mathrm{min};\;\gamma_\mathrm{max}],
\end{equation}
where $\alpha$ is a power of a spectral flux in the optically thin regime: $S_{\nu}\propto\nu^{-\alpha}$. Here $K_{e*}$ is an amplitude of the emitting plasma energy distribution:
\begin{equation}
n_\mathrm{syn*}=\frac{K_{e*}}{2\alpha}\left(\frac{1}{\gamma_\mathrm{min}^{2\alpha}}-\frac{1}{\gamma_\mathrm{max}^{2\alpha}}\right)\approx \frac{K_{e*}}{2\alpha\gamma_\mathrm{min}^{2\alpha}}.
\end{equation}
The relation between the observational frequency $\nu$ and the local properties of a jet at the surface of peak spectral flux (it approximately coincides with the surface with the optical depth equal to unity) is as follows \citep[e.g. Equation (21) in][]{hirotani2005}:
\begin{equation}
\begin{array}{r}
\displaystyle\left(\nu\frac{1+z}{\delta}\right)^{2.5+\alpha}=C_1(\alpha)\frac{1+z}{\delta}\frac{d}{\sin\theta}r_0^2\nu_0\left(\frac{e}{2\pi m c}\right)^{1.5+\alpha}\times \\ \ \\
\displaystyle\times K_{e*}B_*^{1.5+\alpha}.
\end{array}
\label{Eq1}
\end{equation}
Here $m$, $e$, $c$, $r_0$ are the electron mass and charge, the speed of light and the classical electron radius; $\nu_0=c/r_0$; $C_1(0.5)=2.85$ for $\alpha=0.5$ \citep{Gould79}. The local jet width is $d$, the emitting electron number density amplitude is $K_{e*}$ and the local magnetic field in the plasma proper frame is $B_*$. The jet viewing angle is $\theta$, and a Doppler factor for a bulk Lorentz factor $\Gamma=(1-\beta^2)^{-0.5}$ is
\begin{equation}
\delta=\frac{1}{\Gamma(1-\beta\cos\theta)}.
\label{Doppler}
\end{equation}
In order to obtain the dependence of the core position on the observational frequency, we need to set the dependence of of $K_{e*}$, $B_*$, $\Gamma$, $d$ and $\delta$ on the distance along a jet $r$.

The typical frequencies for the core shift measurements are $1.4-43$~GHz \citep{Kov-08, Hada11, Sokolovsky11, Boccardi_2021}, and may be as high as 86 and 230~GHz for M87 \citep{Doeleman_12, Hada_16}. This means that either all the observational frequencies, or the highest ones may probe the jet in its parabolic accelerating domain. 
Thus, we assume here the jet shape is given by (\ref{d_r}), with the fiducial value for power $k=0.5$ for strictly parabolic outflow, and $k\approx 0.4-0.6$ for a quasi-parabolic one.

In general, we assume the power-laws $K_{e*}\propto r^{-k_\mathrm{n}}$ and $B_*\propto r^{-k_\mathrm{b}}$. 
Magnetic field in a jet is decomposed in a poloidal $B_\mathrm{p}$ and toroidal $B_{\varphi}$ components. The toroidal component dominates the major part of a jet $r>R_\mathrm{L}$, as there is a universal relation $B_{\varphi}=B_\mathrm{p}r/R_\mathrm{L}$ for a relativistic flow \citep[see e.g.][]{Lyu09}. However, as the velocity of bulk plasma motion is dominated by a poloidal component, the corresponding relation in the plasma proper frame between $B_\mathrm{p*}\approx B_\mathrm{p}$ and $B_{\varphi*}\approx\sqrt{B_{\varphi}^2-E^2}$ can be different.
We focus on the domain, where jet plasma is still effectively accelerating. In this case either the co-moving magnetic field is dominated by the poloidal component $B_\mathrm{p}^2\gg B_{\varphi}^2-E^2$ or both components in a plasma proper frame are comparable $B_\mathrm{p}^2\sim B_{\varphi}^2-E^2$ \citep{Vlahakis04, Kom09}. In this case we adopt
\begin{equation}
B_*^2\approx B_\mathrm{p}^2.
\end{equation}
To keep the final expression as simple as possible we will not account here for the jet transversal structure. Thus, the conservation of the total magnetic flux defines the dependence of $B_*$ on the distance $r$:
\begin{equation}
B_*(r)=B_{\zeta*}\left(\frac{r_{\zeta}}{r}\right)^{2k},
\label{B}
\end{equation}
where $r_{\zeta}$ is a distance at which we will estimate the magnetic field and emitting plasma number density. It is convenient to use $r_{\zeta}=1$~pc except for the source with the  jet shape break detected at the smaller distances, as NGC 315 \citep{Boccardi_2021, Ricci22}.

The plasma number density obeys the continuity equation and can be written as
\begin{equation}
n_{*1}d_1^2c\Gamma_1=n_{*2}d_2^2c\Gamma_2
\label{cont_eq}
\end{equation}
for the steady state in the accelerating jet.
Here subscripts $1$ and $2$ indicate two different crosscuts perpendicular to the jet axis.
We expect that the jets are still effectively accelerating in the parabolic domain \citep{Nokhrina19, Kov-20, NKP20_r2, Nokhrina2022}, and the plasma bulk motion Lorentz factor
\begin{equation}
\Gamma(d)=\rho\frac{d}{2R_\mathrm{L}}
\label{G}
\end{equation}
grows linearly with the jet radius $d/2$ and depends on the light cylinder radius $R_\mathrm{L}$ \citep{Beskin06, Kom07, Lyu09}. The ideal linear acceleration $\Gamma\propto d$ is valid while the flow is still highly magnetised. In order to account for the slowing down of this Lorentz factor behaviour as the flow approaches the saturation state with the magnetisation equal to unity, we introduce the coefficient $\rho$. Using relation (A13) from \citet{Nokhrina2022} and the Table~1 from \citet{Nokhrina19}, we conclude that $\rho\approx 0.3$ -- $0.55$. Using (\ref{cont_eq}) and (\ref{G}) we obtain the following relation for emitting plasma number density amplitude:
\begin{equation}
K_{e*}=K_{e*\zeta}\left(\frac{r_{\zeta}}{r}\right)^{3k}.
\label{k}
\end{equation}

Substituting (\ref{B}) and (\ref{k}) into (\ref{Eq1}) we obtain the relation for frequency dependent core position
\begin{equation}
r\propto \nu^{-1/k_\mathrm{r}}.
\label{core_shift}
\end{equation}
For the different relations between viewing angle $\theta$ and local Lorentz factor $\Gamma$ in \citet{Ricci22}, we obtain three different values of core shift exponent $k_\mathrm{r}$. For $\theta\approx\Gamma^{-1}$ Doppler factor is approximately constant, and
\begin{equation}
k_\mathrm{r}=\frac{k_\mathrm{n}+(1.5+\alpha)k_\mathrm{b}-k}{2.5+\alpha}.
\label{k2}
\end{equation}
For $k_\mathrm{n}=3k$ and $k_\mathrm{b}=2k$ core shift exponent does not depend on the index of emitting particles distribution $\alpha$: $k_\mathrm{r}=2k$.
If $\theta\ll\Gamma^{-1}$, Doppler factor is proportional to the Lorentz factor $\delta\approx 2\Gamma$, and 
\begin{equation}
k_\mathrm{r}=\frac{k_\mathrm{n}+(1.5+\alpha)k_\mathrm{b}-(2.5+\alpha)k}{2.5+\alpha}.
\end{equation}
Substituting exponents from (\ref{B}) and (\ref{k}) we obtain
\begin{equation}
k_\mathrm{r}=k\frac{3.5+\alpha}{2.5+\alpha}.
\label{k43}
\end{equation}
This expression depends weakly on emitting particle distribution for a reasonable range of $\alpha$ \citep[see, e.g.,][]{Gould79}. For $\alpha=0.5$ this exponent $k_\mathrm{r}=4k/3$.
If $\theta\gg\Gamma^{-1}$ and $\delta\propto\Gamma^{-1}$, one gets
\begin{equation}
k_\mathrm{r}=\frac{k_\mathrm{n}+(1.5+\alpha)k_\mathrm{b}+(0.5+\alpha)k}{2.5+\alpha}.
\end{equation}
For $k_\mathrm{n}=3k$ and $k_\mathrm{b}=2k$
\begin{equation}
k_\mathrm{r}=k\frac{6.5+3\alpha}{2.5+\alpha},
\end{equation}
which is equal to $k_\mathrm{r}=8k/3$ for $\alpha=0.5$ and, again, in general depends on $\alpha$ very weakly. All the above results have been obtained by \citet{Ricci22}, but assuming the dominance of toroidal magnetic field in the collimation and acceleration region.

Obtained $k_\mathrm{r}=2k$ for $\theta\approx\Gamma^{-1}$ for a strictly parabolic power $k=0.5$ provides $k_\mathrm{r}=1$, which coincides with the result for a conical outflow with a constant speed. For a number of sources it is supported by multi frequency measurements for a dozen of sources \citep{Sokolovsky11, Hada11}. However, for different quasi parabolic jet collimation profiles and different viewing angles, $k_\mathrm{r}$ may vary. The core shift exponent measurements for NGC~315 favour $\theta\ll\Gamma^{-1}$ and $k_\mathrm{r}=4k/3$ \citep{Ricci22}.

It is convenient to use the core shift offset \citep{L98}
\begin{equation}
\Omega_{r\nu}=4.8\times 10^{-9}\frac{\Delta r_\mathrm{mas}D_\mathrm{L}}{(1+z)^2}\left(\frac{\nu_1^{1/k_\mathrm{r}}\nu_2^{1/k_\mathrm{r}}}{\nu_2^{1/k_\mathrm{r}}-\nu_1^{1/k_\mathrm{r}}}\right)\;\mathrm{pc\, GHz^{1/k_\mathrm{r}}},
\label{Omega}
\end{equation}
where $\Delta r_\mathrm{mas}$ is a measured core shift in mas for observational frequencies $\nu_1$ and $\nu_2$ for a source at the luminosity distance $D_\mathrm{L}$ and corresponding cosmological red shift $z$. The core-position offset measure depends on $K_{e*\zeta}$ and $B_{*\zeta}$. If the relation between $K_{e*\zeta}$ and $B_{*\zeta}$ is set, measured core shift allows estimating magnetic field and emitting particles number density \citep{L98, hirotani2005, Sullivan09_coreshift, NBKZ15}.

\section{Magnetic field estimates basing on core shift effect}
\label{s:core_shift_B}

The classical equipartition between emitting plasma and magnetic field $K_{e*}\propto B_{*}^2$ cannot hold everywhere anymore due to relations (\ref{B}) and (\ref{k}) being in contrast with the conical approach (see App.~\ref{a:equip}). To relate $K_{e*\zeta}$ and $B_{*\zeta}$ we introduce the local magnetisation --- a ratio of Poynting flux to the plasma bulk motion kinetic energy flux \citep[see, e.g.,][]{NBKZ15}:
\begin{equation}
\sigma=\frac{B^2_{\varphi}c/4\pi}{n_*\Gamma mc^3}.
\end{equation}
Setting $\sigma=1$, which corresponds to $\Gamma=\Gamma_\mathrm{max}/2$, and 
using the universal for relativistic flows relation between toroidal and poloidal magnetic field components 
\citep[e.g.,][]{Vlahakis04, Lyu09, Kom09}
\begin{equation}
\frac{B_{\varphi}}{B_\mathrm{p}}=\frac{d}{2 R_\mathrm{L}},
\end{equation}
we obtain
\begin{equation}
n_{*\sigma=1}=\frac{\Gamma_\mathrm{max}}{8\pi mc^2\rho^2}B_{*\sigma=1}^2.
\end{equation}
In order to relate plasma number density and magnetic field at the given distance $\zeta$ and the point $\sigma=1$, we use (\ref{B}) and (\ref{k}).
This provides the following relation of $n_{*\zeta}$ and $B_{*\zeta}$:
\begin{equation}
n_{*\zeta}=B_{*\zeta}^2\frac{\Gamma_\mathrm{max}}{8\pi m c^2\rho^2}\frac{d_{\zeta}}{d_\mathrm{br}}.
\label{eq_new}
\end{equation}
Assuming that just a fraction $\eta\in(0;\,1]$ of all the plasma emits, we set
\begin{equation}
K_{e*\zeta}=B_{*\zeta}^2\frac{\eta \Gamma_\mathrm{max}}{8\pi m c^2\rho^2}\frac{d_{\zeta}}{d_\mathrm{br}}.
\label{equi}
\end{equation}

With the above relation, we are able to write the final expression for estimating the jet magnetic field using the core shift effect in the parabolic accelerating jet with $\delta\approx\mathrm{const}$ ($k_\mathrm{r}=2k$):
\begin{equation}
\begin{array}{r}
\displaystyle B_{*\zeta}=0.0137\left[\left(\frac{\Omega_{r\nu}}{r_{\zeta}}\right)^{6k}\left(\frac{1+z}{\delta}\right)^2\frac{1}{\sin^{6k-1}\theta}\times\right. \\ \ \\
\displaystyle\times\left.\frac{\rho^2 d_\mathrm{br}}{\eta \Gamma_\mathrm{max}d_{\zeta}^2}\right]^{1/4}\;\mathrm{G}.
\end{array}
\label{Bz}
\end{equation}
In the expressions for  magnetic field $B_{*\zeta}$ all the distances ($r_{\zeta}$, $d_{\zeta}$, $d_\mathrm{br}$ and $R_\mathrm{L}$) are in parsecs.
Compare it with the classical expression for a magnetic field in the plasma proper frame at a distance $1$~pc from the jet base \citep[e.g.,][]{L98, hirotani2005}:
\begin{equation}
B_{*1}=0.025\left(\frac{\Omega_{r\nu}^3(1+z)^2}{\delta^2\varphi\sin^2\theta}\right)^{1/4}\;\mathrm{G},
\label{B1}
\end{equation}
where $\varphi$ is a conical jet intrinsic half-opening angle.

The expression for a magnetic field in a case of $\delta\approx 2\Gamma$ ($k_\mathrm{r}=4k/3$) can be written as:
\begin{equation}
B_{*\zeta}=0.0137\left[\left(\frac{\Omega_{r\nu}}{r_{\zeta}}\right)^{4k}\frac{(1+z)^2}{\sin^{4k-1}\theta}
\frac{R_\mathrm{L}^2\rho^2 d_\mathrm{br}}{\eta \Gamma_\mathrm{max}d_{\zeta}^4}\right]^{1/4}\;\mathrm{G}.
\label{Bz_2}
\end{equation}

Although the case of $k_\mathrm{r}=8k/3$ ($\delta\propto\Gamma^{-1}$) is unlikely, basing on the current observations, we still present the expression for $B$:
\begin{equation}
\begin{array}{r}
\displaystyle 
B_{*\zeta}=0.0097\left[\left(\frac{\Omega_{r\nu}}{r_{\zeta}}\right)^{7k}\frac{(1+z)^2(1-\cos\theta)^2}{\sin^{8k-1}\theta}\times\right. \\ \ \\
\displaystyle\times\left.
\frac{\rho^2 d_\mathrm{br}}{\eta \Gamma_\mathrm{max}R_\mathrm{L}^2d_{\zeta}^{1-k}}\right]^{1/4}\;\mathrm{G}.
\end{array}
\label{Bz_3}
\end{equation}

For the domain of effectively accelerating outflow, the co-moving magnetic field is dominated by the poloidal component \citep{Vlahakis04, Kom09}.
This makes it straightforward estimating both the total magnetic flux contained in a jet
and extrapolating magnetic field down to the gravitational radius scale. 

The total magnetic flux is estimated as
\begin{equation}
\Psi=\pi \left(\frac{d_{\zeta}}{2}\right)^2B_{*\zeta}.
\label{Psi}
\end{equation}
Using the magnetic flux conservation, we obtain the magnetic field amplitude at the gravitational radius scales:
\begin{equation}
B_\mathrm{g}^\mathrm{par}=B_{*\zeta}\left(\frac{r_{\zeta}}{r_\mathrm{g}}\right)^{2k}.
\label{Bg}
\end{equation}

\section{Magnetic field estimates}
\label{s:Bfield}

Below we apply the above relations to the sources with the measured jet shape and core shift. Additionally we use the black hole mass estimate to assess the magnetic field amplitude at the gravitational radius. The needed parameters are summarised in the Tables~\ref{t:param}--\ref{t:param3}.

\begin{table*}
\caption{Parameters of the sources with a measured parabolic jet shape and the core shift.
\label{t:param}}
  \begin{tabular}{cccccccc}
  \hline\hline
 Source & $z$ & $\theta$ & $\theta$  & $M$ & $M$ & {$\delta$} &  $\varphi_\mathrm{app}$ \\
            &     & (deg)    & reference & $\left(\log M_{\odot}\right)$ & reference & &  ($^{\circ}$) \\
 (1)    & (2)               & (3)                   & (4)  & (5) & (6) & (7) & (8) \\ 
\hline
0055$+$300 & 0.017 & 38.0 & \citet{Ricci22} &  9.32 & \citet{Boizelle_2021} & \ldots & $6.9$ \\
0111$+$021 & 0.047 & 5.0 & \citet{Kov-20} & \ldots & \ldots & \ldots & $9.3$ \\
0415$+$379 & 0.049 & 17.4 & \citet{MOJAVE_XIV} & 8.21 & \citet{Trrlb12} & $2.9$ & $7.9$ \\
0430$+$052 & 0.033 & 18.7 & \citet{MOJAVE_XIV} & 8.13 & \citet{WU02} & $5.9$ & $9.6$ \\
1226$+$023 & 0.158 & 3.3 & \citet{MOJAVE_XIV} & 8.41 & \citet{GravCol18} & $17.0$ &  $6.6$ \\
1228$+$126 & 0.004 & 18.0 & \citet{Nakamura+18} & 9.82 & \citet{EHT_I} & 3.2$^*$ & $12.0$ \\
1514$+$004 & 0.052 & 15.0 & \citet{Kov-20} & \ldots & \ldots & \ldots & $7.3$ \\
1637$+$826 & 0.024 & 18.0 & \citet{Kov-20} & 8.78 & \citet{Frrrs99} & \ldots & $7.2$ \\
1807$+$698 & 0.051 & 7.3 & \citet{MOJAVE_XIV} & 8.51 & \citet{WU02} & $1.1$ & $10.6$ \\
2200$+$420 & 0.069 & 8.0 & \citet{MOJAVE_XIV} & 8.23 & \citet{WU02} & $7.3$ & $24.8$ \\
\hline
\end{tabular}
\begin{flushleft} 
{\it Note.} The columns are as follows: 
(1) source name (B1950);
(2) redshift \citep[collected by][]{Kov-20};
(3) viewing angle; 
(4) viewing angle reference;
(5) black hole mass; 
(6) black hole mass reference;
(7) variability Doppler factor from \citet{Hovatta2009}; 
(8) apparent opening angle at $15$~GHz core by \citet{MOJAVE_XIV}. \\
$^*$ For 1228$+$126 Doppler factor is based on the measured at parsec scales velocities by \citet{Mertens16} and the viewing angle.
\end{flushleft}
\end{table*}

\begin{table*}
\caption{Jet geometry parameters of the sources.
\label{t:param2}}
  \begin{tabular}{ccccccc}
  \hline\hline
 Source & Alias & $k$ & $d_\mathrm{break}$ & $r_\mathrm{break}$ & $a_1$ & geometry \\
    & &     & (pc)               & (pc) & ($\mathrm{pc}^{1-k}$) &  reference \\
 (1)    & (2)               & (3)                   & (4)  & (5) & (6) & (7) \\ 
\hline
0055$+$300 & NGC 315 & 0.45  & 0.072 & 0.58 & 0.092 & \citet{Boccardi_2021, Ricci22} \\
0111$+$021 & UGC 00773 & $0.497\pm 0.077$ & 0.28 & 27.31 & $0.179\pm 0.010$ & \citet{Kov-20} \\
0415$+$379 & 3C 111 & $0.468\pm 0.026$ & 0.74 & 29.00 & $0.305\pm 0.011$ & \citet{Kov-20} \\
0430$+$052 & 3C 120 & $0.556\pm 0.070$  & 0.29 & 5.77 & $0.202\pm 0.015$ & \citet{Kov-20} \\
1226$+$023 & 3C 273 & $0.66$ & 3.0 & 274 & 0.223 & \citet{Okino22} \\
1228$+$126 & M 87 & 0.57  & 1.2 & 43.4 & $0.07$ & \citet{Nokhrina19} \\
1514$+$004 & PKS B1514$+$004 & $0.564\pm 0.048$ & 0.34 & 13.1 & $0.171\pm 0.019$ & \citet{Kov-20} \\
1637$+$826 & NGC 6251 & $0.506\pm 0.041$  & 0.16 & 3.3 & $0.155\pm 0.005$ & \citet{Kov-20} \\
1807$+$698 & 3C 371 & $0.388\pm 0.087$  & 0.25 & 12.8 & $0.207\pm 0.016$ & \citet{Kov-20} \\
2200$+$420 & BL Lac &  $0.537\pm 0.057$  & 0.95 & 24.6 & $0.505\pm 0.029$ & \citet{Kov-20} \\
\hline
\end{tabular}
\begin{flushleft} 
{\it Note.} The columns are as follows: 
(1) source name (B1950);
(2) $k$-index upstream the jet shape geometry transition (parabolic flow); 
(3) jet width at the trasnition region; 
(4) jet shape break position from the jet base;
(5) $a_1$ coefficient in a jet shape (parabolic flow); 
(6) reference for a jet geometry results.
\end{flushleft}
\end{table*}

\begin{table*}
\caption{Opacity parameters of the sources and the locations of radio cores and jet shape break points.
\label{t:param3}}
  \begin{tabular}{cccccccc}
  \hline\hline
 Source & $\Delta r$ & $\nu_1$ & $\nu_2$ & reference & $\Omega_{r\nu}$ & $r_{\nu_2}$ & $r_{\nu_1}$ \\
        & (mas) & (GHz) & (GHz) & nuclear opacity & $(\mathrm{pc\,GHz}^{1/k_\mathrm{r}})$ & (pc) & (pc) \\
 (1) & (2) & (3) & (4) & (5) & (6) & (7) & (8) \\ 
\hline
0055$+$300 & 0.420 & 8.4 & 15.4 & \citet{Boccardi_2021} & 8.94 & 0.12 & 0.34 \\
0111$+$021 & 0.137 & 8.1 & 15.4 & \citet{Pushkarev12} & 2.15 & 1.57 & 3.00 \\
0415$+$379 & 0.315 & 8.1 & 15.4 & \citet{Pushkarev12} & 5.62 & 1.01 & 2.01 \\
0430$+$052 & 0.075 & 8.1 & 15.4 & \citet{Pushkarev12} & 0.729 & 0.19 & 0.35 \\
1226$+$023 & 0.036 & 15.4 & 23.7 & \citet{Okino22} & 2.76 & 4.36 & 6.04 \\
1228$+$126 & 0.089 & 8.1 & 15.4 & \citet{Hada11} & 0.113 & 0.028 & 0.051 \\
1514$+$004 & 0.139 & 8.1 & 15.4 & \citet{Pushkarev12} & 2.05 & 0.70 & 1.24 \\
1637$+$826 & 0.210 & 8.1 & 15.4 & \citet{Pushkarev12} & 1.70 & 0.37 & 0.69 \\
1807$+$698 & 0.249 & 8.1 & 15.4 & \citet{Pushkarev12} & 6.44 & 1.49 & 3.41 \\
2200$+$420 & 0.032 & 8.1 & 15.4 & \citet{Pushkarev12} & 0.645 & 0.37 & 0.66 \\
\hline
\end{tabular}
\begin{flushleft} 
{\it Note.} The columns are as follows: 
(1) source name (B1950);
(2) core shift between the low and high frequencies; 
(3) low frequency; 
(4) high frequency; 
(5) reference for a core shift measurement; 
(6) core-position offset $\Omega_{r\nu}$ calculated using columns (2)--(4) and $k_\mathrm{r}$ from Table~(\ref{t:Bfield});
(7) core position from the jet base at the high frequency assuming $k_\mathrm{r}$ from Table~\ref{t:Bfield}; 
(8) the same as column (7), but at the low frequency. 
\end{flushleft}
\end{table*}

To calculate the parameter $\Omega_{r\nu}$ using the expression (\ref{Omega}) we adopt the core shift measurements by \citet{Pushkarev12} between the observational frequencies 15.4 and 8.1 GHz. We also set $k_\mathrm{r}=2k$, where power $k$ in a parabolic regime was measured in \citet{Kov-20} (see Tables~\ref{t:param2} and \ref{t:Bfield}). 
For M~87 jet we use the core shift measurements reported by \citet{Hada11}. 
The core position dependence is
\begin{equation}
r(\nu)=(1.40\pm 0.16)\nu^{-0.94\pm 0.09}+(-0.041\pm 0.012).
\end{equation}
For the homogeneity of the results, we calculate the core shift between the pair of frequencies $8.1$~GHz and $15.4$~GHz (see Table~\ref{t:param3}) using this expression. The obtained by \citet{Hada11} $k_\mathrm{r}$ index assumes the value $1.06^{+0.12}_{-0.09}$, and it is in good agreement with the relation $k_\mathrm{r}=2k$, corresponding to a constant Doppler factor result (\ref{k2}). We set the measured $k_\mathrm{r}=1.06$ for M 87 \citep{Hada11}.
For NGC~315 we use the recent results by \citet{Ricci22}. In particular, multi frequency measurements are consistent with $k_\mathrm{r}=0.57\pm 0.17$ at $22-8$~GHz, and $k_\mathrm{r}=0.92\pm 0.01$ at $8-5$~GHz. This result is in agreement with formula (\ref{k43}) $k_\mathrm{r}=4k/3$, with $4k/3=0.6$ for $k=0.45$ \citep{Boccardi_2021}. We adopt the core shift measurement by \citet{Boccardi_2021} between the frequencies 15.4 and 8.4~GHz. For 3C273 the core shift was measured by \citet{Okino22}. We choose the pair 23.7 and 15.4~GHz as the closest to the frequency pair in other sources. We also adopt $k_\mathrm{r}=2k$, since the authors fitted the core shift with assumed $k_\mathrm{r}=1$. The opacity parameters are listed in the Table~\ref{t:param3}. Please, note, that $\Omega_{r\nu}$ here calculated with $k_\mathrm{r}$, which differs from the unity.
In order to check that the corresponding cores are indeed lying in a quasi parabolic domain, we calculated the positions of cores from a jet base at the high and low frequencies: $r_{\nu_2}$ and $r_{\nu_1}$ --- listed in columns (7) and (8) of the Table~\ref{t:param3}. We see that these cores lie closer to a jet base, than the geometry break point position $r_\mathrm{break}$ (see column (5) in the Table~\ref{t:param2}).

The jet geometry data --- the jet opening parameter $a_1$, power $k$ in a parabolic domain, the jet width at the break $d_\mathrm{br}$ and the position of a break $r_\mathrm{break}$ are presented in the Table~\ref{t:param2}. 

The variability Doppler factors estimates are summarised in the column seven of Table~\ref{t:param} \citep{Hovatta2009}. To estimate the Doppler factor for M~87 jet we employ the maximum speed in the acceleration region measured by \citet{Mertens16} $\Gamma\sim 3$ with the viewing angle $\theta=18^{\circ}$ from \citet{Nakamura+18}. For the sources with unknown Doppler factor we set it as $\delta=1$ basing on relatively large viewing angles of these jets.

In order to compare the magnetic field estimates based on the model of parabolic accelerating jets (\ref{Bz}) with the ones based on conical geometry (\ref{B1}), we use apparent opening angle measurements by \citet{MOJAVE_XIV}. These angles reflect the jet geometry at $15$~GHz core. For M87 jet we use the apparent opening angle $\varphi_\mathrm{app}\approx 5^{\circ}$ reported by \citet{Mertens16} on parsec scales. To obtain the intrinsic half-opening angles, we use the relation
\begin{equation}
\varphi=\varphi_\mathrm{app}\sin\theta/2.
\end{equation}
Viewing angles are collected in \citet{MOJAVE_XIV}. However, for M~87 we use the value $\theta=18^{\circ}$ from \citet{Nakamura+18}, and for NGC~315 --- from \citet{Ricci22}.

We set the parameters $\eta=0.01$, basing on the modelling of synchrotron emission of nonuniform jets \citep{Frolova23}, and $\rho=0.33$ \citep{Nokhrina2022}.

The results are presented in Table~\ref{t:Bfield}. Using the equation~(\ref{Bz}) and the measured parameters, we estimate the magnetic field $B_{*\zeta}^{\mathrm{par}}$ at the distance $\zeta$ (see column(2)) from the jet base for a parabolic accelerating jets. The conical estimate $B_{*\zeta}^{\mathrm{cone}}$ is presented in column (6) to compare the results. We observe that for all the sources except for 3C~273 the magnetic field at the given distance $\zeta$ is about several times smaller that the estimates based on the conical jet geometry. 

Special attention must be given to NGC~315. As the core shift index $k_\mathrm{r}$ at high frequencies clearly points to the case $\theta\ll\Gamma^{-1}$, we should apply the formula (\ref{Bz_2}). We adopt the light cylinder radius $R_\mathrm{L}=4.7\times 10^{-4}$~pc, corresponding to $\Gamma_\mathrm{max}=20$, and calculate magnetic field at 0.2~pc using the relation (\ref{Bz_2}). In this case $B_{*\zeta}$ and $B_\mathrm{g}$ are by the order of magnitude lower, than in the case of the relation (\ref{Bz}).

\begin{table*}
\caption{Estimates of magnetic field.
\label{t:Bfield}}
  \begin{tabular}{cccccccccc}
  \hline\hline
 Source & $\zeta$ & $k_\mathrm{r}$ & $\Gamma_\mathrm{max}$ & $B_{*\zeta}^\mathrm{par}$ & $B_{*\zeta}^\mathrm{cone}$ & $K_{e*\zeta}$ & $\log_{10}\Psi$ & $\Sigma_{\zeta}$ & $B_\mathrm{g}^\mathrm{par}$ \\
  & (pc) & & & (G) & (G) & ($\mathrm{cm}^{-3}$) & ($\mathrm{G\,cm^2}$) & & (G) \\
 (1) & (2)  & (3) & (4) & (5) & (6) & (7) & (8) & (9) & (10) \\ 
\hline
0055$+$300 & 0.2 & 0.57  & 20 & $0.52$ & $0.75$ & $1.2\times 10^{4}$ & 34.3 & $0.11$ & $480$  \\
0055$+$300 & 0.2 & 0.57  & 20 & $0.023$ & $0.75$ & $23$ & 32.9 & $0.11$ & $21$  \\
0111$+$021 & 1.0 & 0.994 & 50 & 0.18 & 0.53 & $1.4\times 10^4$ & 34.2 & 0.12 & $700-3.4\times10^{4*}$  \\
0415$+$379 & 1.0 & 0.936 & 50 & 0.073 & 0.25 & 270 & 34.8 & 0.095 & $4.4\times 10^3$  \\
0430$+$052 & 1.0 & 1.112 & 40 & 0.014 & 0.040 & 13 & 33.7 & 0.075 & $8.0\times 10^3$  \\
1226$+$023 & 1.0 & 1.32 & 30 & 0.16 & 0.13 & $45$ & 33.9 & 2.8 & $4.8\times 10^5$  \\
1228$+$126 & 1.0 & 1.06 & 20 & 0.0082 & 0.012 & 0.35 & 32.5 & 0.95 & 80 \\
1514$+$004 & 1.0 & 1.128 & 50 & 0.11 & 0.27 & $590$ & 34.3 & 0.095 & $1.3\times 10^3-1.1\times 10^{5*}$  \\
1637$+$826 & 1.0 & 1.012 & 40 & 0.059 & 0.18 & $330$ & 34.1 & 0.052 & $2.3\times 10^3$  \\
1807$+$698 & 1.0 & 0.776 & 50 & 0.13 & 0.61 & $1.3\times 10^3$ & 34.5 & 0.060 & $660$  \\
2200$+$420 & 1.0 & 1.074 & 50 & 0.018 & 0.048 & $12.8$ & 34.2 & 0.12 & $5.1\times 10^3$  \\
\hline
\end{tabular}
\begin{flushleft} 
{\it Note.} The columns are as follows: 
(1) source name (B1950);
(2) distance at which we estimate the magnetic field;
(3) power $k_\mathrm{r}$ of a core shift;
(4) assumed maximum Lorentz factor (see details in \citet{NKP20_r2} and \citet{Ricci22});
(5) estimate of magnetic field at distance $\zeta$; we use the formula for a constant Doppler factor (\ref{Bz}) everywhere except the second line of 0055$+$300 (NGC~315), where we use the relation (\ref{Bz_2}); 
(6) magnetic field adopting the conical non-acceleration jet model (\ref{B1}) at the same distance $\zeta$; 
(7) emitting particles number density amplitude $K_{e*\zeta}$ calculated using the relation (\ref{equi}); 
(8) the total magnetic flux in a jet (\ref{Psi}); 
(9) estimate for a standard ratio between magnetic field and emitting plasma energy density, which is assumed to be equal to unity for a conical jet model; 
(10) magnetic field at the gravitational radius scale in parabolic geometry (\ref{Bg}).

$^*$ The magnetic field at the gravitational radius for the sources 0111$+$021 and 1514$+$004 is presented for boundary BH masses $10^{8}-5\times 10^{9}\,M_{\odot}$.
\end{flushleft}
\end{table*}

\section{Results and Discussion}
\label{s:discussion}

First of all, we observe, that the magnetic field estimates within a model of a parabolic accelerating jet is a few times smaller than that for a classical conical model \citep{L98}. The only exception is 3C~273 with  the largest in the sample jet width at the break $d_\mathrm{br}$. 

These differences in a magnetic field calculated within two different models --- conical with a constant velocity and quasi-parabolic accelerating flow --- results from the two effects. The first one is the difference in a dependence of $K_{e*}$ and $B_*$ on the distance $r$. For the ideal parabolic power $k=0.5$ magnetic field $B_*\propto r^{-1}$ as in the Blandford--K\"onigl model. On the other hand, the emitting plasma number density amplitude $K_{e*}\propto r^{-1.5}$ decreases slower than in the flow with a constant Lorentz factor. In a majority of our sources only the first effect plays the role: the core shift parameter $\Omega_{r\nu}$ and the core positions $r_{\nu_2}$ and $r_{\nu_1}$ for $k_\mathrm{r}\approx 1$ are close to their values for the non-accelerating conical outflow. Thus, if we observe the core of accelerating parabolic jet at the same frequency and distance, as in the conical case, than the expression $K_{e*}B_*^{1.5+\alpha}$, which defines the local opacity (see Equation~(\ref{Eq1})) must also be the same. But, as the plasma number density decreases slower with the distance, than the magnetic field amplitude must be lower to balance the larger plasma number density. The second effect is the possible difference in the parabolic and conical core positions for a large enough departures of $k_\mathrm{r}$ from the unity. In this case both effects play the role. If we set in NGC~315 (005$+$300) $k_\mathrm{r}=1$, the core shift offset would be equal to $\Omega_{r\nu}=2.56\;\mathrm{pc\,GHz}$, and the core positions $r_{\nu_1}=0.27$ and $r_{\nu_2}=0.50$~pc. These values are larger than the ones calculated with $k_\mathrm{r}=0.57$, and such a difference in their positions works in the same direction as the first effect, making the magnetic field estimate in a parabolic jet even smaller than in the conical one. In 3C~273 (1226$+$023) for $k_\mathrm{r}=1$ $\Omega_{r\nu}=4.27\;\mathrm{pc\,GHz}$, $r_{\nu_1}=3.13$, $r_{\nu_2}=4.82$~pc.
The cores are situated closer to the jet base than the ones in the model with $k_\mathrm{r}=1.32$. In this case there is an interplay between the effects of slower decrease of $K_{e*}$ with a distance and needed higher amplitudes in both $K_{e*}$ and $B_*$ for the surface with the optical depth equal to unity to be situated farther from the jet origin.

The dominance of a poloidal magnetic field component in the accelerating parabolic jet domain provides a convenient and straightforward way to recalculate the magnetic field onto scales of the order of a gravitational radius. The results are presented in the tenth column in Table~\ref{t:Bfield}. For NGC~315 both the obtained values $B_\mathrm{g}^\mathrm{par}=21$ and $480$~G can be compared with the independent estimate based on an assumption of magnetically arrested disc launching a jet \citep{Ricci22}. The estimate for $k_\mathrm{r}=2k$ is in excellent agreement with $B_\mathrm{MAD}=125-480$~G at the lower limit event horizon radius. On the other hand, the magnetic field, obtained assuming $k_\mathrm{r}=4k/3$, when recalculated to the upper limit magnetosphere radius $10.6\,r_\mathrm{g}$, has the value $2.5$~G, which 
correspond very well to the independent magnetic estimate $5-18$~G at this scale. The $B_\mathrm{g}^\mathrm{par}=80$~G estimate for M~87 is corroborated by the EHT results $B\approx 1-30$~G in the very vicinity of a central BH \citep{EHT_B}. The rest of the sources in the sample have amplitudes of a magnetic field at the gravitational radius of the order of $10^3-10^4$~G, with only 3C~273 reaching a few of $10^5$~G due to relatively large core shift measured by \citet{Okino22}. The estimated total magnetic flux in a jet assumes the expected values of $10^{32}-10^{35}\;\mathrm{G\cdot cm^2}$.

The presented above results depend on the modelling parameters such as the assumed fraction $\eta$ of emitting plasma, the coefficient $\rho$ describing a deviation from the strictly linear acceleration, the maximum possible Lorentz factor $\Gamma_\mathrm{max}$ of a bulk plasma motion and Doppler factor $\delta$ unknown for several sources in the sample. Fraction $\eta=n_\mathrm{syn}/n$ is the most weakly constrained parameter. It is estimated to be from the order of 1\% basing on the particle-in-cell simulations \citep{SironiSpitkovsky2011, SironiSpitkovskyArons2013, SironiSpitkovsky2014} and modelling synchrotron intensity maps from the stratified jets \citep{Frolova23} to the order of 100\% in numerical modelling by \citet{KramerMacDonald2021}. Parameters $\rho$ and $\Gamma_\mathrm{max}$ are usually better constrained. We can estimate $\Gamma_\mathrm{max}$ with the precision up to a factor of a few \citep[see discussion in][]{Nokhrina19}. Both numerical \citep{Chatterjee2019} and analytical \citep{Nokhrina2022} modelling provide about the same accuracy for $\rho\approx 0.2 - 1$. Finally, the Doppler factor is unknown for four of the sources in our sample and assumed to be a unity.

The dependence of estimates (\ref{Bz}) and (\ref{Bz_2}) on the unknown ratio of emitting to full plasma number density and $\Gamma_\mathrm{max}$ is very weak: as the $1/4$ power. This means that even if all the plasma emits ($\eta=1$ instead of $0.01$), magnetic field estimates would increase by the factor of about $3$. This would provide both estimates based on parabolic and conical geometry closer to each other. The same holds for the total magnetic flux $\Psi$ and the magnetic filed at the gravitational radius $B_\mathrm{g}^\mathrm{par}$. Magnetic field depends on the particular values of $\rho$ and $\delta$ as a power $1/2$. Thus, the larger than assumed $\delta=1$ Doppler factors will lower the magnetic field estimate as $1/\sqrt{\delta}\approx 3$ for e.g. $\delta=10$ instead of $1$.
Thus, even the extreme values of these parameters will affect the result for $B$ within a factor of a few. 

Errors in estimates for the magnetic field and emitting plasma number density arise from the observational errors in measured values $\Delta r$, $\theta$, $\delta$, $d_\mathrm{break}$) and from model assumed values $\eta$, $\rho$, $\Gamma_\mathrm{max}$. Typical errors in observational data are of the order of tens percent \citep{Pushkarev12, MOJAVE_XIV, Kov-20}, with the major source being an error in $\Delta r$. The largest expected uncertainty of the model parameters is in the fraction of emitting plasma number density $\eta$. As we discussed above, it may lead to the drop in an estimate of $B_{*\zeta}^\mathrm{par}$ up to 3.2 times. Thus, the results presented in column (5) must be treated as estimates with the accuracy up to, but no more, than a factor of a few.

The estimate $B_{*\zeta}^\mathrm{cone}$ based on conical non-accelerating jet model (\ref{B1}) depends on the apparent intrinsic opening angle $\varphi_\mathrm{app}$, which reflects the local jet width at $r_1=1$~pc. It is changing in the acceleration and collimation zone. We checked that accounting for the true jet width $a_1 r_1^{k}$ provides smaller jet width and at the distance $r_1$, leading to effectively smaller $\varphi$ and, thus, even larger estimate for $B_{*\zeta}^\mathrm{cone}$. However, due to a weak dependence of $B_{*1}$ on $\varphi$ in (\ref{B1}), its value changes at most by a factor of two.

After substitution of the relation (\ref{equi}) between $K_{e*\zeta}$ and $B_{*\zeta}$ into equations (\ref{Bz}) and (\ref{Bz_2}), we obtain the dependence of $K_{e*\zeta}$ on the observed values such as $\delta$, and the implied (modelling) parameters such as $\eta$ and $\rho$.
In case of $k_\mathrm{r}=2k$, the emitting plasma number density depends on the parameters $\eta$, $\Gamma_\mathrm{max}$, $d_\mathrm{br}$, $\rho$ and $\delta$ as
\begin{equation}
K_{e*\zeta}=9.23\left(\frac{\Omega_{r\nu}}{r_\zeta}\right)^{3k}\frac{1+z}{\sin^{(6k-1)/2}\theta}\sqrt{\frac{\eta\Gamma_\mathrm{max}}{d_\mathrm{br}}}\frac{1}{\rho\delta}.
\label{Ke1}
\end{equation}
For $k_\mathrm{r}=4k/3$ the expression for the emitting plasma number density amplitude is
\begin{equation}
K_{e*\zeta}=9.23\left(\frac{\Omega_{r\nu}}{r_\zeta}\right)^{2k}\frac{1}{d_\zeta}\frac{1+z}{\sin^{(4k-1)/2}\theta}\sqrt{\frac{\eta\Gamma_\mathrm{max}}{d_\mathrm{br}}}\frac{R_\mathrm{L}}{\rho}.
\label{Ke2}
\end{equation}
We see that $K_{e*\zeta}$ estimate would grow by the factor of $10$ if all the plasma emits ($\eta=100$ instead of $1$). The dependence of $K_{e*\zeta}$ on the Doppler factor is even stronger. It we take for four sources, for example, $\delta=10$ instead of $1$, the estimate in emitting plasma number density will fall by the same factor $10$. It is worth noting that the sources with assumed $\delta=1$ have the largest estimates for emitting plasma number density. Using fiducial $\delta=10$ would decrease these values much closer to the ones for the sources with observational estimates of Doppler factor.

The new relation between $K_{e*}$ and $B_*$ presented in this work $K_{e*}\propto B_*^2d$ is very important since it differs from usually accepted equipartition
\begin{equation}
\frac{B_*^2}{8\pi}=\int_{\gamma_\mathrm{min}}^{\gamma_\mathrm{max}}mc^2\gamma dn_\mathrm{syn*}\approx K_{e*}mc^2\Lambda.
\label{class_equi}
\end{equation}
Here $\Lambda=\ln(\gamma_\mathrm{max}/\gamma_\mathrm{min})$ for $\alpha=0.5$, and we used in our calculations $\Lambda=10$. If we assume this classical equipartition regime, than the core shift offset $\Omega_{r\nu}$ should depend on the distance along a jet, or, correspondingly, on the particular frequency pair. 
See the details in Appendix~\ref{a:equip}. 

The relation (\ref{equi}) means that while a jet is still accelerating, the ratio $K_{e*}/B$ grows proportional to the jet width until the acceleration ceases. This is due to dependence of particle number density in a plasma proper frame on the local Lorentz factor. As soon as the flow reaches almost constant $\Gamma$, the additional dependence on a jet width, connected with $\Gamma$, vanishes, and the relation between $K_{e*}$ and $B_*^2$ assumes naturally the classical form (\ref{class_equi}). In order to get a feeling of difference from the classical equipartition, we calculate the values
\begin{equation}
\Sigma_{\zeta}=\frac{B_*^2}{8\pi mc^2K_{e*}\Lambda}
\end{equation}
(presented in the column 9 in Table~\ref{t:Bfield}), which should be equal to unity for a classical equipartition regime.

Equations (\ref{Bz}), (\ref{Bz_2}) and (\ref{Bz_3}) are similar to the Equations (13), (14) and (15) by \citet{Ricci22} with one important distinction. While the assumed dependence of both $B_*$ and $K_{e*}$ on $r$ are the same in both works, the difference comes from the euqipartition assumptions, and it affects the dependence of $B_*$ on the $d_{\zeta}$. \citet{Ricci22} assumed $K_{e*\zeta}\propto B_{*\zeta}^2$ everywhere, and here we show that it is principal to set the relation $K_{e*\zeta}\propto B_{*\zeta}^2d_\zeta$ for the consistency of the results (See Appendix~A). This means that the Equations (13)--(15) by \citet{Ricci22} can be applied for the comparable jet widths at the cores $r_{\nu_i}$ and at the point where we estimate the magnetic field $\zeta$. Only in this case the obtained results are self-consistent. This condition holds for the for NGC~315 in \citet{Ricci22}, since $d_\mathrm{break}=0.072$, $d_{\nu_1}=0.057$ and $d_{\nu_1}=0.035$~pc. Extrapolating our magnetic field estimate from thee distance $\zeta=0.2$~pc to $\zeta=0.58$~pc, we get $B_*=0.2$~G, which is very close to the result $0.18$~G by \citet{Ricci22}.

There is a zone along a jet, where the $k$-index changes more or less smoothly from quasi-parabolic value to quasi-conical. Observational data provides a change in $k$ on the short length scales \citep[see e.g.][]{Kov-20, Boccardi_2021}, while the analytical modelling provides a wider range for such a change \citep{Kov-20}. Here we assume the constant $k$ in a quasi parabolic domain, since the radio cores lie much closer to the jet origin than the geometry break position (see Tables~{\ref{t:param2} and \ref{t:param3}}). However, if the core at the lower frequency lies in the transition domain with the $k$-index being in between the parabolic and conical values, it will boost the magnetic field estimate closer to the results within the conical non-accelerating model.

\section{Summary}
\label{s:summary}

This work is dedicated to the modification of a core shift method developed by \citet{L98} for parabolic accelerating jets in the region of the observed radio cores. We apply our results to the sample of sources with the detected jet shape break. Our results are summarised as follows.

1. We use the poloidal component of a magnetic field as the dominant component $B_*$ in the plasma proper frame. This is true for a still accelerating jet \citep{Vlahakis04, Kom09}. We account for a changing particle number density in a plasma proper frame $K_{e*}$ as the flow Lorentz factor grows. As a result, the relation between $K_{e*}$ and $B_*^2$ deviates from the classical equipartition. The values of the core shift offset $\Omega_{r\nu}$ for different pairs of frequencies do not show any significant trend with $r$. It supports our using the relation (\ref{equi}) between $K_{e*}$ and $B_*^2$.

2. We show that for a jet boundary geometry given by a power-law $d\propto r^k$ the core shift index in the parabolic accelerating jet may assume values $k_\mathrm{r}=2k$, $k_\mathrm{r}=4k/3$ and $k_\mathrm{r}=8k/3$ for $\theta\ll\Gamma^{-1}$ (Doppler factor $\delta\approx 2\Gamma$), $\theta\sim\Gamma^{-1}$ (Doppler factor $\delta\approx\mathrm{const}$) and $\theta\gg\Gamma^{-1}$ (Doppler factor $\delta\propto\Gamma^{-1}$) respectively. Thus, for quasi parabolic jets, we should expect $k_\mathrm{r}\approx 0.8-1.2$ and $k_\mathrm{r}\approx 0.53-0.8$ for the most of the sources. Both values are observed \citep[e.g.,][]{Hada11, Sokolovsky11, Porth_etal11, kutkin_etal14}, and we explain these values by the different viewing angles with respect to the emission opening angle $\sim\Gamma^{-1}$. In fact, this means, that multi frequency observations of radio cores coupled with the kinematics can provide one more instrument to constrain the jet viewing angle.

3. We obtained the formulas to estimate magnetic field $B_*$ (\ref{Bz}) and (\ref{Bz_2}), and the emitting particles number density amplitude $K_{e*}$ (\ref{equi}) using the measurements of a core shift effect and a jet geometry. We show that $B_*$ can be straightforwardly used to calculate the magnetic field at the gravitational radius scales and the total magnetic flux in a jet. This is due to the dominance of a poloidal component in the accelerating jet region. 

4. We have estimated a magnetic field in a jet $B_*$ and on a gravitational radius $B_\mathrm{g}$, a particle number density $K_{e*}$ and a total magnetic flux $\Psi$ for a sample of ten sources. The magnetic field at 1~pc distance along a jet is a few times smaller than the classical estimates. For NGC~315 we probed both equations (\ref{Bz}) and (\ref{Bz_2}), although the measured by \citet{Ricci22} $k_\mathrm{r}=0.57$ strongly points to the case of $\theta\ll\Gamma^{-1}$. For M~87 and NGC~315 $B_\mathrm{g}$ can be compared with the independent estimates by different methods, and the correspondence is very good.

\section*{Acknowledgements}

We thank the anonymous referee for suggestions
which helped to improve the paper.
This study has been supported
by the Russian Science Foundation: project 20-72-10078, https://rscf.ru/project/20-72-10078/.
This research made use of the data from the MOJAVE database\footnote{\url{http://www.physics.purdue.edu/MOJAVE/}} which is maintained by the MOJAVE team \citep{MOJAVE_XV}.
This research made use of NASA's Astrophysics Data System.

\section*{Data availability}

There is no new data associated with the results presented in the paper. All the previously published data has the proper references.
Intereferometric raw data of the project code BK134 is available from the VLBA archive\footnote{\url{https://data.nrao.edu}}.

\bibliographystyle{mnras}
\bibliography{nee1}

\appendix

\section{Possible relations between plasma number density and magnetic field}
\label{a:equip}

\begin{figure*}
    \centering
    \includegraphics[width=0.45\linewidth]{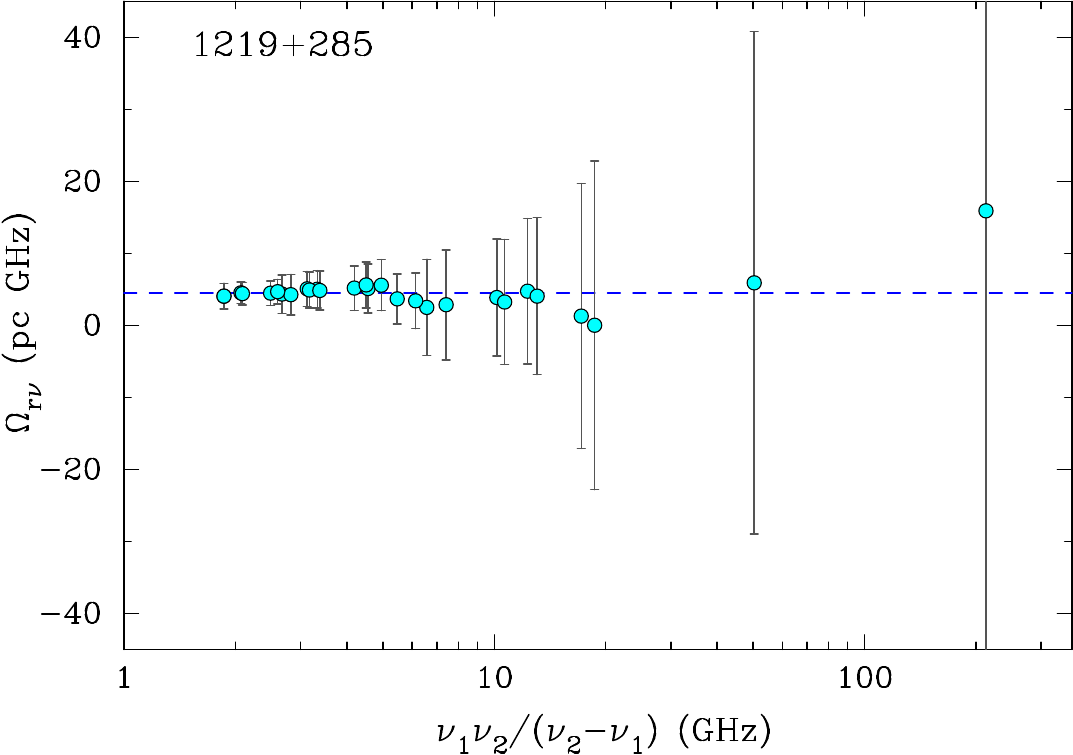} \hspace{0.1cm}
    \includegraphics[width=0.45\linewidth]{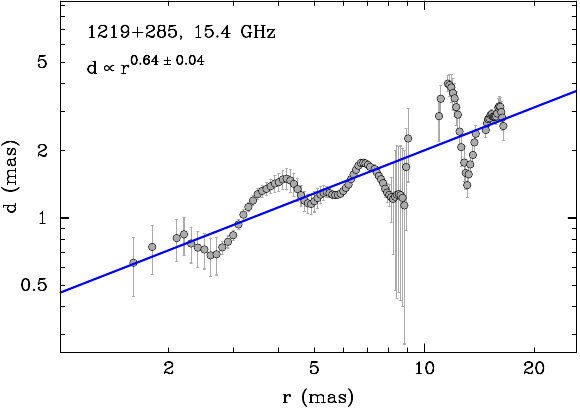}   
    \caption{Left: Dependence of $\Omega_{r\nu}$ on a frequency parameter 
    $\nu_1\nu_2/(\nu_2-\nu_1)$ for BL Lacertae object 1219+285 constructed 
    from VLBI observations at 1.4, 1.7, 2.3, 2.4, 4.6, 5.0, 8.1, 8.4 and 15.4~GHz. 
    Dashed line denotes the median $\Omega_{r\nu}$.  Smaller x-axis values 
    correspond to wider frequency pairs and vice versa.  
    Right: Evolution of the jet width as a function of the VLBI core separation 
    at 15.4~GHz showing a quasi-parabolic streamline. Solid blue line represents 
    the fitted dependency.}
    \label{f:omega}
\end{figure*}

Suppose there is some kind of relation between the emitting particles number density amplitude $k_{e*}$ and the magnetic field $B_*$ in the region of the observed radio cores. The usually used one is the equipartition between energy density of emitting plasma and magnetic field $k_{e*}\propto B_*^2$ \citep{Burbidge56, BlandfordKoenigl1979, L98}. Suppose also, that the the multi frequency observations are indeed consistent with the power-law dependence (\ref{core_shift}), which means a constant (at least for the implied distances along a jet) $\Omega_{r\nu}$ \citep[e.g.,][]{Sokolovsky11, Hada11}. 
Let us discuss the possible relations between the emitting plasma number density and the magnetic field in the plasma proper frame, that ensure $\Omega_{r\nu}$ being approximately constant for different frequency pairs. The latter, in turn, means constant value at different distances $r_{\zeta}$.

Substituting relations $K_{e*}\propto r^{-k_\mathrm{n}}$, $B_*\propto r^{-k_\mathrm{b}}$ and (\ref{d_r}) into Equation~(\ref{Eq1}), we obtain the following:
\begin{equation}
\Omega_{r\nu}^{p}=Cr_\zeta^{k_\mathrm{n}+k_\mathrm{b}(1.5+\alpha)}K_{e*\zeta}B_{*\zeta}^{1.5+\alpha},
\end{equation}
where the power $p$ and the coefficient $C$ depend on assumed Doppler factor dependence on the Lorentz factor.

Case 1. Approximately constant Doppler factor $\delta$.
\begin{equation}
p=k_\mathrm{n}+k_\mathrm{b}(1.5+\alpha)-k,
\end{equation}
\begin{equation}
\begin{array}{r}
\displaystyle C=3.086\times 10^{-9(0.5+\alpha)}C_1(\alpha)r_0^2\nu_0\left(\frac{e}{2\pi m c}\right)^{1.5+\alpha}\times\\ \ \\
\displaystyle\times \left(\frac{\delta}{1+z}\right)^{1.5+\alpha}a_1\sin^{p-1}\theta,
\end{array}
\end{equation}
where
$r_0$, $\nu_0$, $e$ are in CGS units, and $a_1$ -- in $\mathrm{pc}^{1-k}$.

Case 2. Doppler factor $\delta\approx 2\Gamma$.
\begin{equation}
p=k_\mathrm{n}+k_\mathrm{b}(1.5+\alpha)-k(2.5+\alpha),
\end{equation}
\begin{equation}
\begin{array}{r}
\displaystyle C=3.086\times 10^{-9(0.5+\alpha)}C_1(\alpha)r_0^2\nu_0\left(\frac{e}{2\pi m c}\right)^{1.5+\alpha}\times\\ \ \\
\displaystyle\times \frac{\rho^{1.5+\alpha}a_1^{2.5+\alpha}\sin^{p-1}\theta}{(1+z)^{1.5+\alpha}R_{\mathrm{L}}^{1.5+\alpha}}.
\end{array}
\end{equation}
where
$R_{\mathrm{L}}$ is in pc.

Case 3. Doppler factor $\delta\propto\Gamma^{-1}$.
\begin{equation}
p=k_\mathrm{n}+k_\mathrm{b}(1.5+\alpha)+k(0.5+\alpha),
\end{equation}
\begin{equation}
\begin{array}{r}
\displaystyle C=3.086\times 10^{-9(0.5+\alpha)}C_1(\alpha)r_0^2\nu_0\left(\frac{e}{2\pi m c}\right)^{1.5+\alpha}\times\\ \ \\
\displaystyle\times \frac{\sin^{p-1}\theta}{a_1^{0.5+\alpha}}\left(\frac{2R_\mathrm{L}}{\rho(1-\cos\theta)(1+z)}\right)^{1.5+\alpha}.
\end{array}
\end{equation}
where
$R_{\mathrm{L}}$ in in pc.

We see, that $\Omega_{r\nu}^p$ is proportional to the expression
\begin{equation}
K_{e*\zeta}B_{*\zeta}^{1.5+\alpha}r_{\zeta}^{k_\mathrm{n}+k_\mathrm{b}(1.5+\alpha)},
\label{Omega_const}
\end{equation}
which, in turn, we do not expect to depend on the particular distance along a jet $\zeta$.

Suppose, the relation $K_{e*}\propto B_*^2$ holds together with $K_{e*}\propto r^{-k_\mathrm{n}}$ and $B_*\propto r^{-k_\mathrm{b}}$. If the expression (\ref{Omega_const}) does not depend on distance $r_\zeta$, the following equality must be true:
\begin{equation}
k_\mathrm{n}=2k_\mathrm{b}.
\end{equation}
This equality is independent of the particular index of plasma energy distribution $\alpha$. It holds, for an instance, for classical approach \citep{L98}
with $k_\mathrm{n}=2$ and $k_\mathrm{b}=1$. However, it does not hold for the assumptions (\ref{B}) and (\ref{k}). The self-consistent dependence of $\Omega_{r\nu}$ on $K_{e*\zeta}$, $B_{*\zeta}$ and $r_{\zeta}$ exists only for a certain relations between emitting plasma number density and magnetic field.
Suppose, one can relate the number density amplitude with the magnetic field as
\begin{equation}
K_{e*\zeta}\propto B_{*\zeta}^{l}r_{\zeta}^{m}.
\end{equation}
In order the combination (\ref{Omega_const}) does not depend on $\zeta$, the following relation between exponents $l$ and $m$ must hold:
\begin{equation}
k_\mathrm{n}+m=k_\mathrm{b}l.
\label{cond-1}
\end{equation}
In expression (\ref{eq_new}) $l=2$ and $m=k$, so the condition (\ref{cond-1}) is indeed fulfilled, and it has an underlying clear physical meaning, as described in section~\ref{s:Bfield}. This relation holds for any dependence of the Doppler factor $\delta$ on $\Gamma$, i.e. for any relation between the viewing angle $\theta$ and Lorentz factor $\Gamma$, as long as expressions (\ref{B}) and (\ref{k}) hold. 

To examine whether there is significant trends of $\Omega_{r\nu}$ with $r$, we analysed the very long baseline interferometry (VLBI) observations of 20 pre-selected sources carried out during four 24h sessions between March and June 2007 (project code BK134) aimed on core shift analysis \citep{Sokolovsky11}.  The observations were performed simultaneously at nine frequencies, ranging from 1.4 to 15.4~GHz, thus forming both narrow and wide frequency pairs and enabling us to probe different spatial scales of the inner jet regions of the target sources.  In Figure~\ref{f:omega} (left), we show that $\Omega_{r\nu}$ is quite stable for the relatively close, $z=0.107$ \citep{Paiano17}, i.e. the luminosity distance $D_L=465$~Mpc, BL Lacertae object 1219+285 (W Comae), assuming $k_r=1$. We note that the inner jet shape of the source is quasi-parabolic (Figure~\ref{f:omega}, right), as inferred from the observations at the highest frequency of 15.4~GHz by analysing profiles transverse to the total intensity ridgeline of the outflow that we constructed following the procedure described in \citet{MOJAVE_XIV} in detail. Among the other 19 sources characterised by different inner jet shapes, such as quasi-parabolic and quasi-conical, a similar flat frequency dependence on $\Omega_{r\nu}$ is observed.

\bsp    
\label{lastpage}

\end{document}